\def\z{\tilde{\omega}}
 \def\zz{\tilde{x}}

\documentclass[prb,preprint]{revtex4-1} % style for Physical Review B and AJP are similar

\usepackage{amsmath} % need for subequations

\usepackage{graphicx} % for figures
\usepackage{amsmath}    % need for subequations
\usepackage{graphicx}   % for figures
\usepackage{draftcopy}
\usepackage{hyperref}
\hypersetup{
    pdfstartview={FitH},
   pdfpagemode={None}
   }
\newcommand{\be}{\begin{equation}}
\newcommand{\ee}{\end{equation}}

\begin{document}

\title{An elementary renormalization-group approach to the Generalized Central Limit Theorem and Extreme Value Distributions}

\author{Ariel Amir}

 \affiliation{John A. Paulsson School of Engineering and Applied Sciences, Harvard University, Cambridge, Massachusetts 02138, USA}
 \email{arielamir@seas.harvard.edu}   %optional

\date{\today}
\begin{abstract}
The Generalized Central Limit Theorem is a remarkable generalization of the Central Limit Theorem, showing that the sum of a large number of independent, identically-distributed (i.i.d) random variables with infinite variance may converge under appropriate scaling to a distribution belonging to a special family known as L\'{e}vy stable distributions. Similarly, the \emph{maximum} of i.i.d. variables may converge to a distribution belonging to one of three universality classes (Gumbel, Weibull and Fr\'{e}chet). Here, we rederive these known results following a mathematically non-rigorous yet highly transparent renormalization-group-like approach that captures both of these universal results following a nearly identical procedure.
\end{abstract}

\maketitle

\section{Introduction}

Consider some distribution $P(x)$ from which we draw independent random variables $x_1, x_2, ... , x_n$. If the distribution has a finite standard deviation $\sigma$ and mean $\langle x \rangle$, we can define:
\be  \xi \equiv \frac{\sum_{i=1}^N (x_i-\langle x \rangle)}{\sigma \sqrt{N}}, \label{CLT} \ee
and the Central Limit Theorem (CLT) tells us that the distribution of $\xi$, $p(\xi)$, approaches a Gaussian with vanishing mean and a standard deviation of 1 as $N \to \infty$. What happens when $P(x)$ does \emph{not} have a finite variance? Or a finite mean? Perhaps surprisingly, in this case the Generalized Central Limit Theoreom (GCLT) tells us that the limiting distribution belongs to a particular family ( L\'{e}vy stable distributions), of which the Gaussian distribution is a proud member albeit \emph{e pluribus unum}. Moreover, the familiar $\sqrt{N}$ scaling of the above equation does not hold in general, and its substitute will generally sensitively depend on the form of the \emph{tail} of the distribution.

The results are particularly intriguing in the case of heavy-tailed distributions where the \emph{mean} diverges. In that case the sum of $N$ variables will be dominated by rare events, regardless of how large $N$ is! Fig. \ref{sum_}c shows one such example, where a running sum of variables drawn from a distribution whose tail falls off as $p(x)\sim 1/x^{3/2}$ was used. The code which generates this figure is remarkably simple, and included in the Appendix.
The underlying reason for this peculiar result is that for distributions with a power-law tail $p(x) \propto 1/x^{1+\mu}$, with $\mu<1$, the distributions of both the \emph{sum} and \emph{maximum} of the $N$ variables scale in the same way with $N$, namely as $N^{1/\mu}$ -- dramatically different from the $\sqrt{N}$ scaling we are used to from the CLT. The distribution of the \emph{maximum} is known as the \emph{Extreme Value Distribution} or EVD (since for large $N$ it inherently deals with rare, atypical events among the $N$ i.i.d variables). Surprisingly, also for this quantity universal statements can be made, and when appropriately scaled this random variable also converges to one of three universality classes -- depending on the nature of the tails of the original distribution from which the i.i.d variables are drawn.

Here, we will provide a straightforward derivation of these results. Although compact and elementary, to the best of our knowledge it has not been utilized previously, and is distinct (and simpler) than other renormalization-group approaches to the GCLT and to EVD. The derivation will not be mathematically rigorous -- in fact, we will not even specify the precise conditions for the theorems to hold, or make precise statements about convergence. In this sense the derivation may be considered as ``exact but not rigorous", targeting a physics rather than mathematics audience (Ref. \onlinecite{feller}, for example, provides a rigorous treatment of many of the results derived in this paper).

Throughout, we will assume sufficiently smooth probability distributions (what mathematicians refer to as probability density functions), potentially with a power-law tail such that the variance or mean may diverge (known as a ``fat" or ``heavy" tail).
%
%\section{Probability distribution of sums}
%\index{Characteristic function}
%An important quantity to define is the \emph{characteristic function} of the distribution:
%\be \varphi(\omega) = \int_{-\infty}^{\infty} f(t) e^{i \omega t} dt .\ee
%Clearly $\varphi(0)=1$, and we have $|\varphi(\omega)|<1$ for $\omega \neq 0$.  Defining:
%
%\be X = x_1 + x_2 + ... + x_n, \ee
%we can write:
%\be \varphi_{sum}( \omega) = \int_{-\infty}^{\infty} ... \int_{-\infty}^{\infty} p(x_1)p(x_2)..p(x_n) dx_1 \; dx_2 \; ... \; dx_n \; \delta (x_1 + x_2 + ... + x_n - X) e^{i\omega X} dX, \ee
%
%which upon doing the integration over $X$ leads to the \emph{product} of the characteristic functions.

\subsection{Example: Cauchy distribution}

Consider the following distribution, known as the Cauchy distribution:
\be p(x)=\frac{1}{\gamma \pi (1+(\frac{x}{\gamma})^2)}. \label{cauchy} \ee
Its characteristic function $\varphi(\omega) \equiv \int_{-\infty}^{\infty} p(x) e^{i \omega x} dx $ is:
\be \varphi(\omega)=e^{-\gamma |\omega|}. \ee
Thus the characteristic function of a sum of $N$ such variables is:
\be \varphi_N(\omega)=e^{-N \gamma |\omega|}, \ee
and taking the inverse Fourier transform we find that the distribution of the sum, $p_N(x)$, is \emph{also} a Cauchy distribution:
\be p_N(x)=\frac{1}{N \gamma \pi (1+(\frac{x}{N\gamma})^2)}.\ee

Thus, the sum does not converge to a Gaussian, but rather retains its Lorentzian form. Moreover, it is interesting to note that the scaling form
governing the width of the Lorentzian evolves with $N$ in a different way than the Gaussian scenario: while in the latter the variance increases linearly with $N$ hence the width increases as $\sqrt{N}$, here the scaling factor is linear in $N$. This remarkable property is in fact useful for certain computer science algorithms \cite{CS}.

\section{Self-similarity of running sums}

%is generated by remarkably simple code, the essence of which is included in the Appendix. Basically,

\begin{figure}[h!]
(a) \includegraphics[width=7 cm]{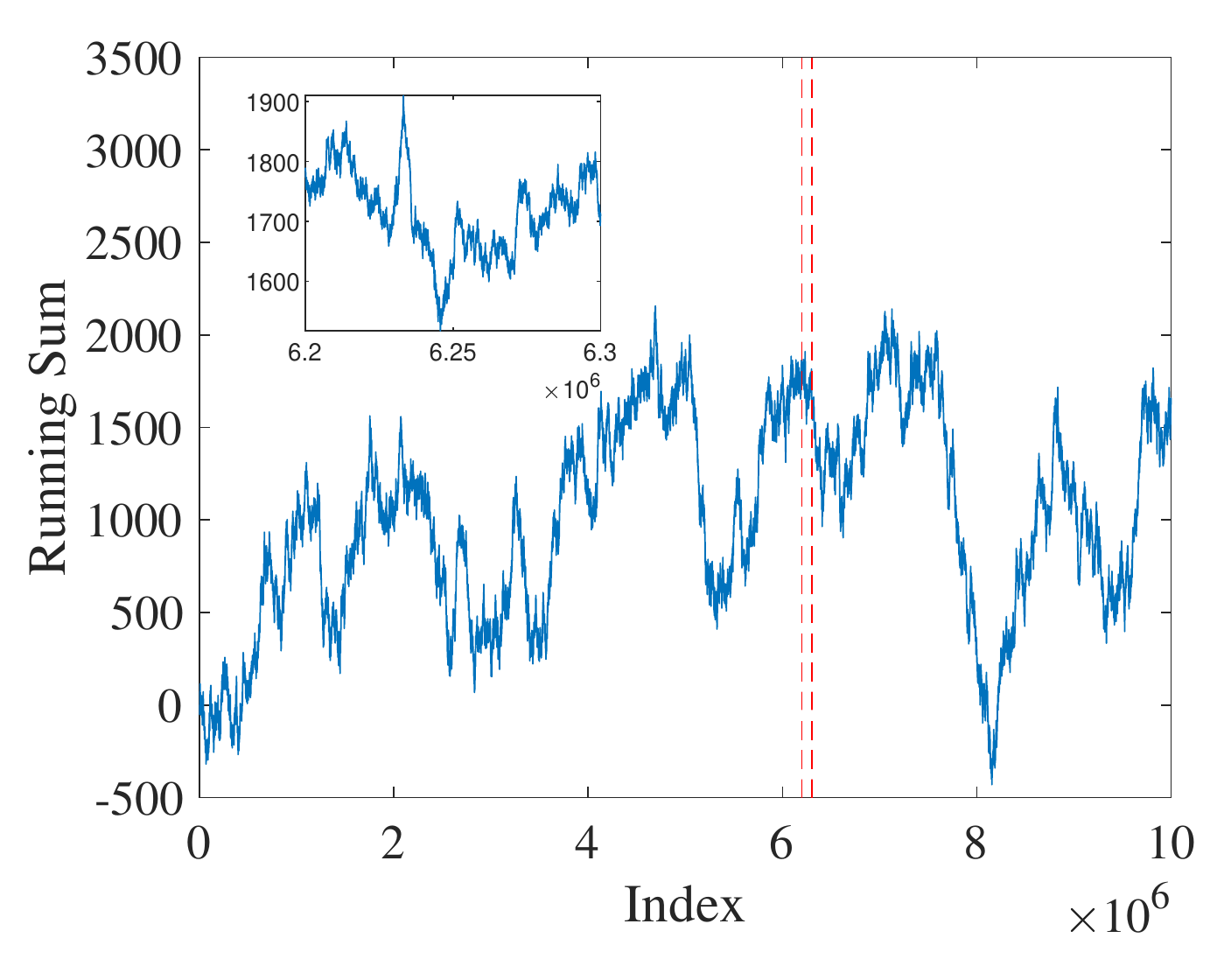} \\
 (b) \includegraphics[width=7 cm]{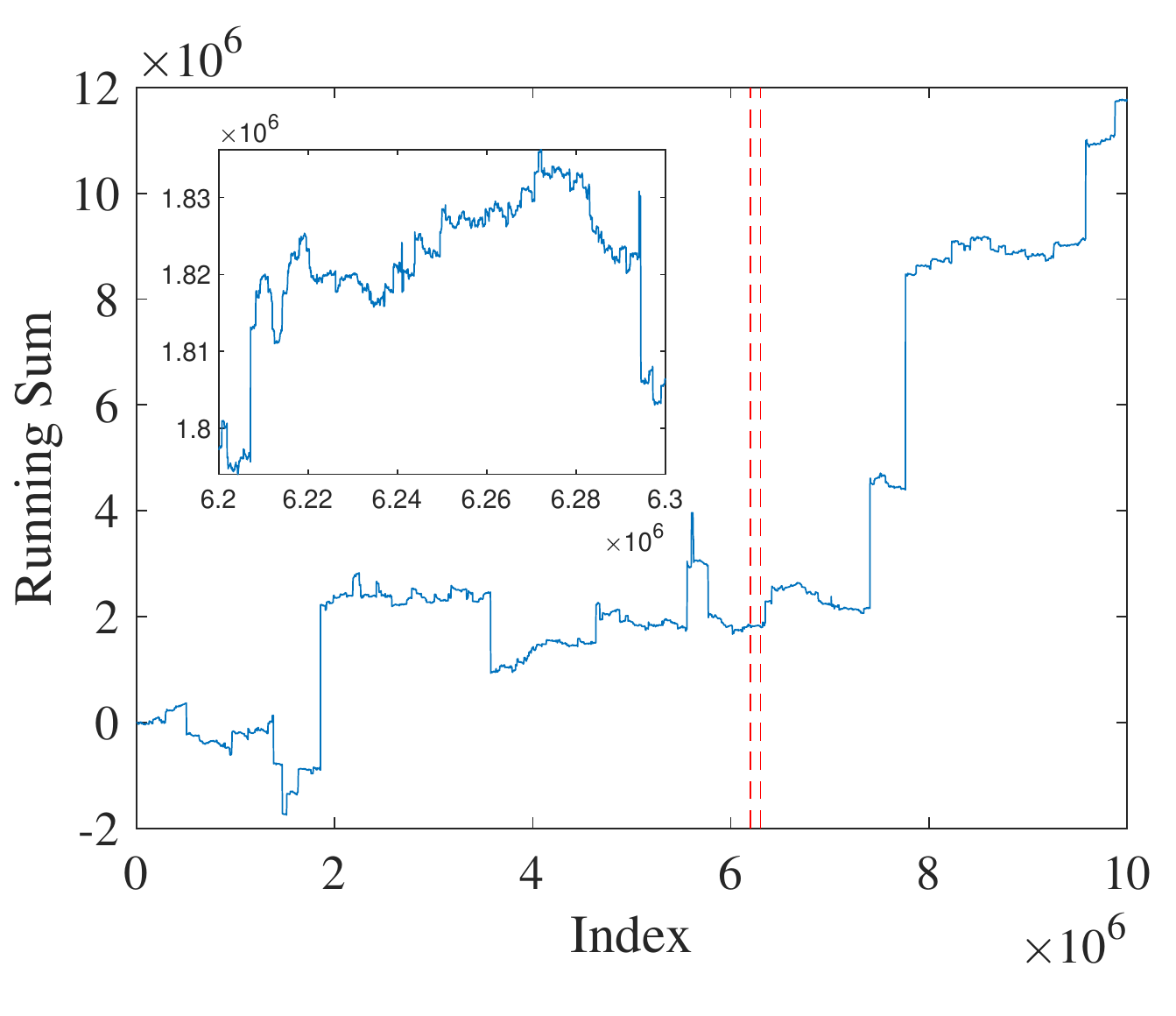} \\
 (c) \includegraphics[width=7 cm]{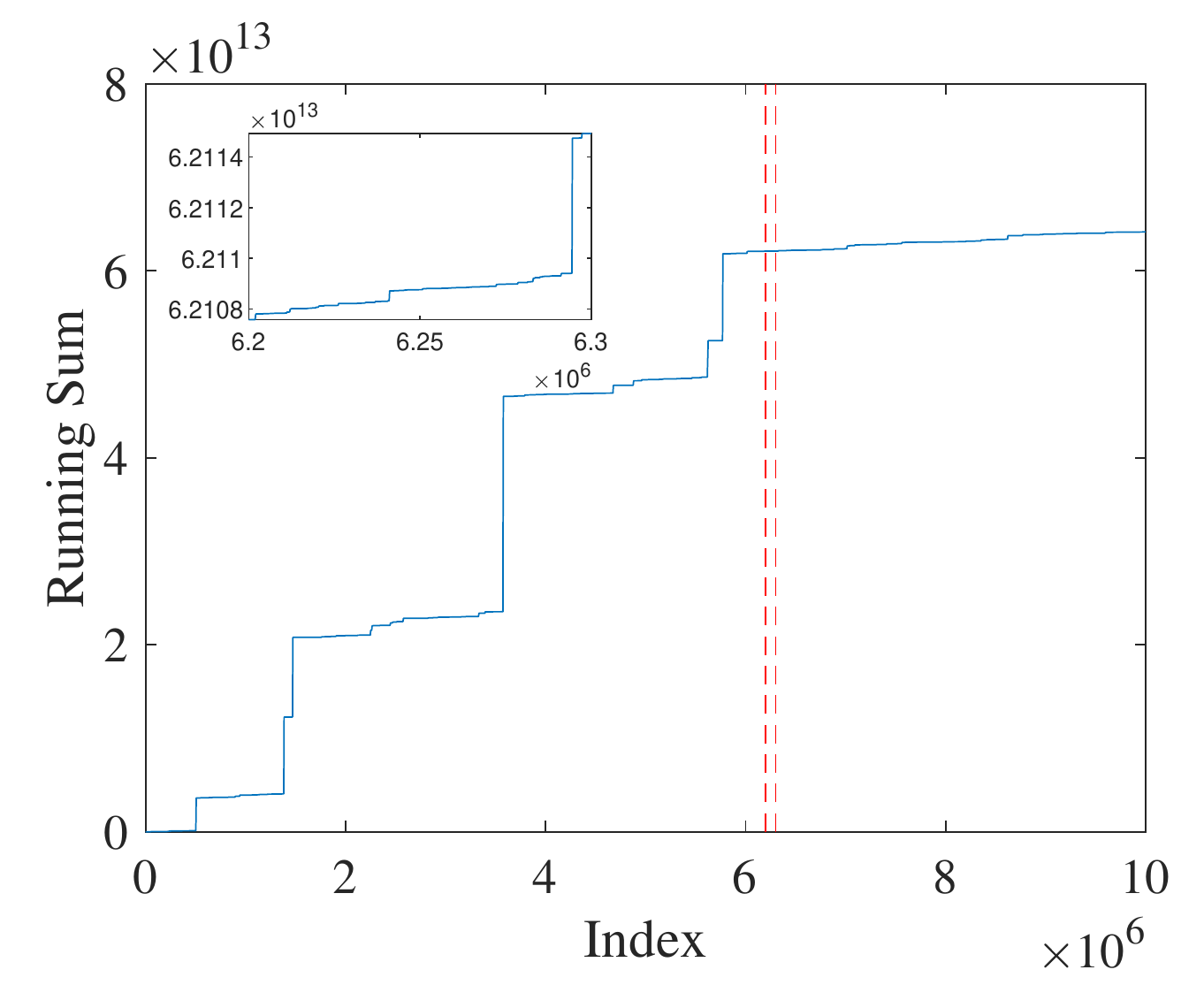} \\
 \caption{Running sum of independent, identically-distributed variables drawn from three distributions: Gaussian (top), Cauchy (middle) and a distribution with positive support and a power-law tail $1/x^{3/2}$ (bottom). See Appendix for details of the code. The insets illustrate the self-similar nature of the running sum, zooming into the small region of the original plot between the two vertical, dashed lines.}
\label{sum_}
\end{figure}

To generate Fig. \ref{sum_}, we generate a set of i.i.d variables from a given distribution (Gaussian, Cauchy and a heavy-tailed distribution whose tail falls off as $1/x^{3/2}$). For each long sequence of random variables, the running sum is plotted. For the Gaussian case (or any case where the variance is finite), the result is the familiar process of diffusion : the variance increases linearly with ``time" (i.e., the index of the running sum). For Fig. \ref{sum_}a the mean vanishes, hence the running sum follows this random walk behavior. If we were to repeat this simulation many times, the result of the running sum at time $N$, a random variable of course, is such that when scaled by $1/\sqrt{N}$ it would follow a normal distribution with variance 1, as noted in Eq. (\ref{CLT}). Another important property is that ``zooming" into the running sum (see the figure inset) looks identical to the original figure -- as long as we don't zoom in too far as to reveal the granularity of the data. Fig. \ref{sum_}b shows the same analysis for the Cauchy distribution of Eq. (\ref{cauchy}). As we have seen, now the scaling is linear in $N$. Nevertheless, zooming into the data still retains its Cauchy statistics. The mathematical procedure we will shortly follow to find \emph{all} L\'{e}vy stable distributions will rely on this self-similarity. Indeed, assume that a sum of variables from some distribution converges --upon appropriate linear scaling --  to some L\'{e}vy stable distribution. Zooming further ``out" corresponds to generating sums of L\'{e}vy stable variables, hence it retains its statistics. A dramatic manifestation of this is shown in Fig. \ref{sum_}c. The initial distribution is \emph{not} a L\'{e}vy stable, but happens to have a very fat tail, possessing infinite mean and variance. The statistics of the running sum converges to a L\'{e}vy stable distribution -- in this cases fortuitously expressible in closed form, corresponding to the L\'{e}vy distribution we will discuss in Eq. (\ref{levy_dist}). Importantly, zooming into the running sum still retains its statistics, which in this case happens to manifest large jumps associated with the phenomenon of L\'{e}vy flights, which will be elucidated by our later analysis. Due to the self-similar nature, zooming into what seems to be flat regions in the graph shows that their statistical structure is the same, and they also exhibit these massive jumps. We also note that this renormalization-group idea has been utilized in the context of the ``conventional" CLT in Ref. \onlinecite{sethna} (Exercise 12.11). Notably, Ref. \onlinecite{jona2001renormalization} discusses the deep connections between the CLT and RG approaches, and emphasizes the notion and relevance of ``Self-similar random fields", highly related to the self-similarity discussed here that forms the basis of our RG-inspired analysis. It should be emphasized that our approach is not an RG one \emph{par excellence} and appears to be simpler than other RG approaches previously utilized in the context of the GCLT \cite{calvo2010generalized}.

\newpage

%generated by running_sum_all

\newpage
\section{Generalized Central Limit Theorem}
\label{general_CLT}
\index{Generalized Central Limit Theorem}
We will look for distributions which are \emph{stable}: this means that if we add two (or more) variables drawn from this distribution, the distribution of the sum will retain the same shape -- i.e., it will be identical up to a potential shift and scaling, by some yet undetermined factors. If the sum of a large number of variables drawn from \emph{any} distribution converges to a distribution with a well defined shape, it must be such a stable distribution. The family of such distributions is known as \emph{L\'{e}vy stable}.

We shall now use an RG (renormalization group) approach to find the general form of such distributions, which will turn out to have a simple representation in Fourier rather than real space -- essentially because the characteristic function is the natural object to deal with here.

The essence of the approach relies on the fact that if we sum a large number of variables, and the sum converges to a stable distribution, then by definition taking a sum involving, say, twice the number of variables, will \emph{also} converge to the same distribution -- up to a potential shift and rescaling. This is illustrated visually in Fig. \ref{sum_} by plotting the running sum of independently and identically distributed variables.

Defining the partial sums by $s_n$, the general (linear) scaling one may consider is:

\be \xi_n = \frac{s_n-b_n}{a_n}. \label{scaling_form} \ee

Here, $a_n$ determines the width of the distribution, and $b_n$ is a shift. If the distribution has a finite mean it seems plausible that we should center it by choosing $b_n = \langle x \rangle n$. We will show that this is indeed the case, and that if its mean is infinite we can set $b_n=0$.

The scaling we are seeking is of the form of Eq. (\ref{scaling_form}), and our hope is that if the distribution of $\xi_n$ is $p_{\xi_n}(x)$, then:

\be \lim_{n \to \infty} p_{\xi_n}(x) = p(x) \label {limit_dist}\ee
exists, i.e., the scaled sum converges to some distribution $p(x)$, which is not necessarily Gaussian (or symmetric).

Let us denote the characteristic function of the scaled variable by $\varphi(\omega)$ (assumed to be approximately independent of $n$ for large $n$). Consider the variable $y_n=\xi_n a_n$. Its distribution, $p_{y_n}$ is:

\be p_{y_n}(y_n) = \frac{1}{a_n} p(y_n/a_n),\ee
with $p$ the limiting distribution of Eq. (\ref{limit_dist}) (and the factor $\frac{1}{a_n}$ arising from the Jacobian of the transformation). The characteristic function of the variable $y_n$ is:

\be \varphi_{y_n}(\omega) = \varphi(a_n \omega). \ee
Consider next the distribution of the sum $s_n$. We have $s_n = y_n + b_n$, and its distribution, $p_{s_n}$, is:

\be p_{s_n}(s_n) = p_{y_n}(s_n-b_n).\ee

Shifting a distribution by $b_n$ implies multiplying the characteristic function by $e^{i \omega b_n}$.
Therefore the characteristic function of the sum is:

\be \varphi_{s_n}(\omega) = e^{i b_n \omega} \varphi_{y_n}(\omega) = e^{i b_n \omega} \varphi(a_n \omega). \label{char_funct} \ee

This form will be the basis for the rest of the derivation, where we emphasize our assumption that the characteristic function $\varphi$ is $n$-independent.

Consider $N = n \cdot m$, where $n,m$ are two large numbers. The important insight is to realize that one may compute $s_N$ in two ways: as the sum of $N$ of the original variables, or as the sum of $m$ variables, each one being the sum of $n$ of the original variables. The characteristic function of the sum of $n$ variables drawn from the original distribution is given by Eq. (\ref{char_funct}). If we take a sum of $m$ variables drawn from \emph{that} distribution (i.e., the one corresponding to the sums of $n$'s), then its characteristic function will be on the one hand:

\be  \varphi_{s_N}(\omega) = e^{i m b_n \omega} (\varphi(a_n \omega))^m, \ee
and on the other hand it is the distribution of $n \cdot m = N$ variables drawn from the original distribution, and hence does not depend on $n$ or $m$ separately but only on their product $N$.  Therefore, assuming that $n$ is sufficiently large such that we may treat it as a continuous variable, we have:

\be \frac{\partial}{\partial n}e^{i \frac{N}{n} b_n \omega + \frac{N}{n}log[\varphi(a_n\omega)]}=0. \ee

Defining $d_n \equiv \frac{b_n}{n}$, we find:

\be \Rightarrow i N \omega \frac{\partial d_n}{\partial n} -\frac{N}{n^2}\log(\varphi)+\frac{N}{n}\frac{\varphi'}{\varphi}\frac{\partial a_n}{\partial n}\omega=0 .\ee

\be \Rightarrow \frac{\varphi'(a_n \omega)\omega}{\varphi(a_n \omega) }=\frac{\log(\varphi(a_n \omega))}{n\frac{\partial a_n}{\partial n}}-i  \omega \frac{\partial d_n}{\partial n} \frac{n}{\frac{\partial a_n}{\partial n}} .\ee

Multiplying both sides by $a_n$ and defining $\z \equiv a_n \omega$, we find that:

\be \frac{\varphi'(\z)\z}{\varphi(\z)}-\log(\varphi(\z)) \frac{a_n}{n\frac{\partial a_n}{\partial n}}+i\z \frac{\partial d_n}{\partial n} \frac{ n}{\frac{\partial a_n}{\partial n}} =0. \label{scale} \ee

Since this equation should hold (with the same function $\varphi(\z)$) as we vary $n$, we expect that $\frac{a_n}{n\frac{\partial a_n}{\partial n}}$ and $\frac{\partial d_n}{\partial n} \frac{ n}{\frac{\partial a_n}{\partial n}}$ should be nearly independent of $n$ for large values of $n$. The equation for $\varphi(\z)$ then takes the following mathematical structure:

\be \frac{\varphi'}{\varphi} -\frac{C_1 \log(\varphi(\z))}{\z}=i C_2 , \label{RG_eq_}\ee

with $C_1, C_2$ constants. We may rewrite it using $u(\z) \equiv \log(\varphi(\z))$ as:

\be u' - \frac{C_1}{\z} u = i C_2. \ee

Therefore the ODE for $u(\z)$ is \emph{linear}, and we may follow the general approach of solving the homogenous equation and guessing a particular solution for the inhomogeneous equation (alternatively, we may note that this is a Cauchy--Euler differential equation and can readily be converted into an ODE equation with constant coefficients using the transformation $t \equiv \log(x)$).
Setting $C_2=0$, we can write the homogenous equation in the form:

\be (\log[u])' = \frac{C_1}{\z}. \ee

%\be \frac{\varphi'}{\varphi \log(\varphi)} = \frac{\log(\varphi)'}{\log(\varphi)} = [\log(\log(\varphi))]'=\frac{C_1}{z} .\ee
Upon integrating we find:
\be  \log(u)=C_1 \log|\z|+const \Rightarrow u =A|\z|^{C_1}. \label{RG} \ee

This is the general solution to the homogenous equation. Note that due to the $1/\z$ term in the equation, this solution is valid for $\z>0$ (for a particular choice of $A$), as well as for $\z<0$ -- but the parameter $A$ can (and will) change between this two regimes, as we shall later see.

Guessing a particular solution to the inhomogeneous equation in the form $u = D \z$, leads to:

\be D - C_1 D = i C_2 \implies D = \frac{i C_2}{1-C_1}. \label{RG2} \ee

As long as $C_1 \neq 1$, we found a solution. In terms of $\varphi(\z)$, the general solution to Eq. (\ref{RG_eq_}) thus takes the form:

\be \varphi(\z) = e^{A|\z|^{C_1} + D \z}. \ee

In the case $C_1=1$, we can guess a solution of the form $u =  D \z \log(\z)$, and find:

\be D\log(\z) + D -D\log(\z) = i C_2, \label{RG3} \ee

hence we have a solution when $D = i C_2$, which leads to the following form for $\varphi$:

\be \varphi(\z) = e^{A|\z|^{C_1} + D \z log(\z)}. \ee

Going back to Eq. (\ref{scale}), we can also get the approximate scaling for the coefficients:

\be \frac{a_n}{n\frac{\partial a_n}{\partial n}} \approx C_1 \implies C_1 \frac{\partial \log(a_n)}{\partial n} \approx 1/n. \ee

This implies that:

\be \log(a_n) \approx \frac{1}{C_1} \log(n) + constant \implies a_n \propto n^{1/C_1}. \ee

Similarly:

\be \frac{\partial d_n}{\partial n} \frac{n}{\frac{\partial a_n}{\partial n}}=C_2.\ee

Hence:
\be \frac{\partial d_n}{\partial n} \propto n^{1/C_1-2}.\ee

Therefore:

\be d_n = C_3 n^{1/C_1-1} + C_4 \implies b_n = C_3 n^{1/C_1} + C_4 n. \ee

The first term will become a constant when we divide by the term $a_n \propto n^{1/C_1}$ of Eq. (\ref{scaling_form}), leading to a simple shift of the resulting distribution. Upon dividing by the term $a_n$, the second term will vanish for large $n$ when $C_1<1$. We shall soon see that the case $C_1>1$ corresponds to the case of a variable with finite mean, in which case the $C_4 n$ term will be associated with centering of the scaled variable by subtracting their mean, as in the standard CLT.

\vspace{0.5 cm}
\noindent\fbox{%
    \parbox{\textwidth}{%
        \textbf{A word of caution.}
The constraint imposed by the RG approach is insufficient in pinning down the scaling factor $a_n$ precisely. Really, all we know is that $ \lim_{n \to \infty} \frac{a_n}{n\frac{\partial a_n}{\partial n}}$ should tend to a constant. In the above, we solved the ODE resulting from equating this term to a constant, but it is easy to see that modulating this power-law by, e.g., logarithmic corrections (or powers thereof) would also satisfy the RG requirement. Similarly care should be taken in interpreting the power-law scaling of the coefficients $b_n$, as well as their counterparts in the ``Extreme Value Distributions" later on.
}
}

\vspace{0.5 cm}

\section{General formula for the characteristic function}

According to Eqs. (\ref{RG}) and (\ref{RG2}), the general formula for the characteristic function of $p(\xi_n)$ for $C_1 \neq 1$ is:

\be \varphi(\omega) = e^{A|\omega|^{C_1} + D \omega}. \label{general_form} \ee

the $D$ term is associated with a trivial shift of the distribution (related to the linear scaling of $b_n$) and can be eliminated. We will therefore not consider it in the following. The case of $C_1=1$ will be considered in the next section.

%It is important to note that the value of A can change at $\omega=0$, where the function is non-analytic. Therefore $\varphi$ assumes the form:

The requirement that the inverse Fourier transform of $\varphi$ is a probability distribution imposes that $\varphi(-\omega)=\varphi^*(\omega)$.
Therefore the characteristic function takes the form:

\be \varphi = \begin{cases} e^{A \omega^{C_1}} &\mbox{ } \omega > 0 \\
e^{A^* |\omega|^{C_1}} &\mbox{ } \omega < 0. \end{cases} \ee
(As noted previously, the value of $A$ in Eq. (\ref{general_form}) was indeed ``allowed" to change at $\omega=0$).
%$C_1=C_2\equiv \mu$, $A_1=A_2^*$, and $\varphi$ take the form

This may be rewritten as:
\be \varphi=e^{-a |\omega|^\mu[1-i \beta sign(\omega) tan(\frac{\pi \mu}{2})]}, \label{levy_dist}\ee
where clearly $\mu = C_1$. The asymmetry of the distribution is determined by $\beta$. For this representation of $\varphi$, we will now show that $-1 \leq \beta \leq 1$, that $\beta=1$ ($\beta=-1)$ corresponds to a distribution with positive (negative) support, and $\beta=0$ corresponds to a symmetric distribution. %This formula will not be valid for $\mu=1$, a case that we will discuss in the next section.

%If we take the Fourier transform of $\varphi$ we obtain that the tails of the distribution are:
%
%$$ p(x) = \begin{cases} \frac{A_+}{x^{1+\alpha}} &\mbox{ } x \rightarrow \infty \\
%\frac{A_-}{x^{1+\alpha}} &\mbox{ } x \rightarrow -\infty \end{cases} $$
%
%with
%
%$$\beta = \frac{A_+-A_-}{A_++A_-}. $$

Consider $p(x)$ which decays, for $x>x^*$, as

\be p(x)=\frac{A_+}{x^{1+\mu}}, \ee

with $0 < \mu< 1$. We will explicitly assume that the function decays sufficiently fast for $x \to -\infty$, and later generalize to the case of both a right and left power-law tail. We shall now find the form of the Fourier transform of $p(x)$ near the origin in terms of the tail of the distribution. For small, positive $\omega$ we find:

\be \Phi(\omega) \equiv \int_{x^*}^{\infty} \frac{A_+}{x^{1+\mu}} e^{i\omega x} dx = A_+\int_{{x^*} \omega}^{\infty} \frac{e^{im}}{m^{1+\mu}}dm \frac{\omega^{1+\mu}}{\omega},\label{PhiOmega} \ee

where we substituted $m= \omega x$. Evaluating the integral on the RHS by parts we obtain:

 \be \Phi(\omega) = A_+ \left[ -\omega^{\mu} \frac{m^{-\mu}}{\mu}e^{im} \big|_{{x^*} \omega}^{\infty}\right] + A_+ \omega^{\mu} \int_{{x^*} \omega}^{\infty} \frac{im^{-\mu}}{\mu} e^{im}dm . \label{omega_int}\ee

 For $\mu<1$, we may approximate the integral by replacing the lower limit of integration by 0, to find:
\be \Phi(\omega)  \approx A_+ \frac{{x^*}^{-\mu}}{\mu}e^{i{x^*} \omega} +  A_+ \frac{\omega^{\mu}}{\mu} i \int_0^{\infty}\frac{e^{im}}{m^{\mu}}dm, \label{lower_cutoff}\ee

and $\int_0^{\infty}\frac{e^{im}}{m^{\mu}}dm=i\Gamma(1-\mu)e^{-i\frac{\pi}{2}\mu}$ (this can easily be evaluated using contour integration). Thus, for small $\omega$ we have:

\be  \Phi(\omega) \approx C_0 -C_+ \omega^{\mu} ,\ee
with $C_0, C_+$ constants. $\Phi(\omega)$ is not the characteristic function, since the lower limit of the integration in Eq. (\ref{PhiOmega}) is $x^*$. However, due to our assumption that the left tail decays fast, we can bound the rest of the integral to be at most linear in $\omega$ for small $\omega$ (up to a constant). Similar results hold if the left tail is power-law albeit with a larger power than the right tail. Therefore the characteristic function near the origin is approximated by:

\be  \varphi(\omega) \approx 1-C_+ \omega^{\mu} , \label{C_right}\ee with $C_+ = A_+ \frac{\Gamma(1-\mu)e^{-i\frac{\pi}{2}\mu}}{\mu}$.
Thus we have:
\be  \frac{\textrm{Im}(C_+)}{\textrm{Re}(C_+)}=-tan({\frac{\pi}{2}\mu}) \Rightarrow C_+ = a [ 1-i \cdot tan({\frac{\pi}{2}\mu})] ,\label{C_define}\ee
with $a$ a real coefficient.
This corresponds to $\beta=1$ in our previous representation of Eq. (\ref{levy_dist}). If we similarly look at a distribution with a left tail, a similar analysis leads to the same form of Eq. (\ref{C_right}) albeit with $C_- = a[1+i \cdot tan({\frac{\pi}{2}\mu})]$ (and $a$ real), corresponding to $\beta=-1$ in Eq. (\ref{levy_dist}). In the general case where both $A_+$ and $A_-$ exist, we obtain the expression

\be \varphi(\omega) \approx 1-\tilde{C} |\omega|^\mu(A_+ e^{-i \mu \frac{\pi}{2}} + A_- e^{i \mu \frac{\pi}{2}}) , \ee

with $\tilde{C} \equiv \frac{\Gamma(1-\mu)}{\mu}$.

We can write this as $\varphi(\omega) \approx 1- C |\omega|^\mu$, where now we have:

\be \frac{\textrm{Im}(C)}{\textrm{Re}(C)}=\frac{-sin(\frac{\pi}{2}\mu)}{cos(\frac{\pi}{2}\mu)} \left(\frac{A_+-A_-}{A_++A_-}\right) = -tan(\frac{\pi}{2}\mu)\beta, \ee

with $\beta$ defined as:

\be \beta = \frac{A_+-A_-}{A_++A_-}. \ee

This clarifies the notation of Eq. (\ref{levy_dist}), and why $\beta$ is restricted to the range $[-1,1]$.

\vspace{0.5 cm}

\noindent\fbox{%
    \parbox{\textwidth}{%
        \textbf{A tail of tales -- and black swans.} \index{Black swans}
It is interesting to note that unlike the case of finite variance, here the limiting distribution depends only on $A_+$ and $A_-$: the tails of the original distribution. The behavior is only dominated by these tails -- even if the power-law behavior only sets in at large values of $x$!
This also brings us to concept of a ``black swan":  scenarios in which rare events -- the probability of which is determined by the tails of the distribution -- yet may have dramatic consequences. Here, such events dominate the sums. For a popular discussion of black swans and their significance, see Ref. \onlinecite{taleb}.
}
}
\vspace{0.5 cm}

What about the case where $1<\mu<2$?
Following the same logic -- with an additional integration by parts - we find that the form of Eq. (\ref{levy_dist})(with $|\beta|\leq 1$) is still intact also for $1<\mu<2$. Note that the linear term will drop out due to the shift of Eq. (\ref{scaling_form}).
It is also worth mentioning that the asymmetry term vanishes as $\mu \to 2$: in the case of finite variance we always obtain a symmetric Gaussian, i.e., we become insensitive to the asymmetry of the original distribution.

\textbf{Special Cases}

\textbf{$\mu=1/2$, $\beta=1$: L\'{e}vy distribution}

Consider the L\'{e}vy distribution:
\be p(x) = \sqrt{\frac{C}{2\pi}}\frac{e^{-\frac{C}{2x}}}{(x)^{3/2}} \;\;\;\; (x\geq 0) \label{levy_dist_2}\ee

The Fourier transform of $p(x)$ for $\omega>0 $ is

\be \varphi(\omega)=e^{-\sqrt{-2i C \omega}},\ee

which indeed correspond to $\tan(\frac{\pi}{2}\frac{1}{2})=1 \rightarrow \beta=1$.

\textbf{$\mu=1$: Cauchy distribution and more}

The case $\mu=1$, $\beta=0$ corresponds to the Cauchy distribution. In the general case $\mu=1$ and $\beta \neq 0$, we have seen that the general form of the characteristic function is, according to Eqs. (\ref{RG}) and (\ref{RG3}):

\be \varphi(\omega) = e^{A|\omega|^{C_1} + D \omega \log(\omega)}. \ee

Repeating the logic we used before to establish the coefficients $D$ for $\omega>0$ and $\omega<0$, based on the power-law tails of the distributions (which in this case fall off like $1/x^2$) leads to:

\be \varphi(\omega) = e^{- |C \omega| [1-i\beta Sign(\omega)\phi]}; \;\;\;\; \phi=-\frac{2}{\pi}log|\omega|.\label{mu_1}\ee

This is the only exception to the form of Eq. (\ref{levy_dist}). It remains to be shown why $\beta=1$ corresponds to the case of a strictly positive distribution (which would thus justify the $\frac{2}{\pi}$ factor in the definition of $\phi$). To see this, note that the logic following up to Eq. (\ref{omega_int}) is still intact for the case $\mu=1$. However, we can no longer replace the lower limit of integration by 0 in Eq. (\ref{lower_cutoff}). The real part of the integral can be evaluated by parts, leading to a  $-\log(\omega)$ divergence. The imaginary part of the integral does not suffer from such a divergence and can be approximated by replacing the lower limit of integration with 0. Using
$\int_0^{\infty} \frac{sin(x)}{x} dx =\pi/2$, we find that:

\be \Phi(\omega) \approx c[\pi/2 + i \log(\omega)], \ee
with $c$ a real number, leading to the form of Eq. (\ref{mu_1}).

\section{RG Approach for Extreme Value Distributions}
\index{Extreme Value Distributions}
Consider the maximum of $n$ variables drawn from some distribution $p(x)$, characterized by a cumulative distribution $C(x)$ (i.e., $C(x)$ is the probability for the variable to be smaller than $x$). It vanishes for $x \to -\infty$ and approaches 1 as $x \to \infty$.
We will now find the behavior of the maximum for large $n$, that will turn out to also follow universal statistics -- much like in the case of the GCLT -- that depend on the tails of $p(x)$. This was discovered by Fisher and Tippett, motivated by an attempt to characterize the distribution of strengths of cotton fibers \cite{fisher}. Our approach will be reminiscent (yet distinct) from that of Fisher and Tippett, and will in fact closely follow the RG-type approach we used for deriving the L\'{e}vy stable distributions, albeit with the \emph{cumulative} distribution function replacing the role of the \emph{characteristic} function -- for reasons that will shortly become clear. Note that other works in the literature also use an RG approach to study this problem, but in a rather different way (e.g., Refs. \onlinecite{RG1, RG2, sethna_PRL, sethna, gyorgyi2008finite, gyorgyi2010renormalization}). While here the derivation only relies on the fact that taking the maximum through different procedures should lead to the same result (in the spirit of RG approaches), these works use a ``traditional" renormalization group approach.

\vspace{0.5 cm}
\noindent\fbox{%
    \parbox{\textwidth}{%
        \textbf{Extreme values.}
        We will be interested in the maximum (or minimum) of a \emph{large} number of variables. By nature, this (rare) random event is an outlier -- the largest or smallest over many trials (assumed here to be independent). Indeed, the results are often applied to problems where the extreme events matter -- what should be the height of a dam? What is the chance of observing an earth-quake or tsunami of a given magnitude? For these reasons insurance companies are likely to be interested in this topic.
        }%
}

\vspace{0.5 cm}

To begin, we define:

\be X_n \equiv \text{max}(x_1,x_2,...,x_n), \ee
where $x_1,...,x_n$ are again i.i.d. variables.
Since we have:

\be Prob(X_n < x) = Prob(x_1 < x) Prob(x_2 <x)... Prob(x_n <x) = C^n(x),\ee
it is natural to work with the cumulative distribution when dealing with extreme value statistics, akin to the role which the characteristic function played in the previous section. Clearly, it is easy to convert the question of the \emph{minimum} of $n$ variables to one related to the maximum, if we define $\tilde{p}(x)=p(-x)$.

Before proceeding to the general analysis, which will yield three distinct universality classes (corresponding to the Gumbel, Weibull and Fr\'{e}chet distributions), we will first exemplify the behavior of each class on a particular example.

\subsection {Example I: the Gumbel distribution}

Consider the distribution:

\be p(x) = e^{-x}. \ee

Its cumulative is:

\be C(x) = 1-e^{-x}. \ee

The cumulative distribution for the maximum of $n$ variables is therefore:

\be G(x) = (1-e^{-x})^n \approx e^{-n e^{-x}} = e^{-e^{-(x-x_0)}}, \label{approx} \ee

with $x_0 \equiv \log(n)$.

This is an example of the \emph{Gumbel distribution}. The general form of its cumulative is:

\be G(x)= e^{ -e^{- (a x +b)}}.\ee

Taking the derivative of Eq. (\ref{approx}) to find the probability distribution for the maximum, we find:
\be p_n(x) = e^{-e^{-(x-x_0)}} e^{-(x-x_0)}, \ee
(where the $n$ dependence enters only via $x_0$).
Denoting $l \equiv e^{-(x-x_0)}$, we have:

\be p(x)=e^{-l} l , \ee and taking the derivative with respect to $l$ we find that the distribution is peaked at $x=\log(n)$. It is easy to see that its width is of order unity. We can now revisit the approximation we made in Eq. (\ref{approx}), and check its validity.

Rewriting $(1-x/n)^n = e^{n \log(1-x/n)}$ and Taylor expanding the exponent to second order, we find that the approximation
\be (1-x/n)^n \approx e^{-x}, \ee
is valid under the condition $x \ll \sqrt{n}$. In our case, this implies:

\be e^{-x} n \ll \sqrt{n} . \label{approx2} \ee

At the peak of the distribution ($x=\log(n)$), we have $e^{-x} n =1$, and the approximation is clearly valid there for $n \gg 1$. From Eq. (\ref{approx2}) we see that the approximation we used would break down when we take $x$ to sufficiently smaller than $\log(n)$. Defining $x = \log(n) - \delta x$, we see that the value of $\delta x$ for which the approximation fails obeys $e^{-\delta x} = O(\sqrt{n})$, hence $\delta x = O(\log(\sqrt{n}))$. Since as we saw earlier the width of the distribution is of order unity, this implies that for large $n$ the Gumbel distribution would approximate the exact solution well, failing only sufficiently far in the (inner) tail where the probability distribution is vanishingly small. However, a note of caution is in place: the logarithmic dependence we found signals a very slow converge to the limiting form. This is also true in the case where the distribution $p(x)$ is Gaussian, as was already noted in Fisher and Tippett's original work \cite{fisher}.

%We will soon show that for the maximum of a large number of variables drawn from \emph{any} distributions with a sufficiently fast decaying tail, the Gumbel distribution arises.

\subsection {Example II: the Weibull distribution}

Consider the \emph{minimum} of the same distribution we had in the previous example.
The same logic would give us that:

\be Prob[min(x_1,..x_n)>\xi] = Prob[x_1>\xi] Prob[x_2>\xi]..Prob[x_n>\xi]= e^{-\xi n}. \ee

This is an example of the Weibull distribution, which occurs when the variable is bounded (e.g: in this case the variable is never negative).

As we shall see below, the general case, for the case of a maximum of $n$ variables with distribution bounded by $x^*$, would be:

\be G(x) = \begin{cases} e^{-a \left(x^*-x\right)^{1/\alpha}}, &   x \leq x^* \\ 0, &   x > x^* \end{cases} \ee

In this case, the behavior of the original distribution $p(x)$ near the cutoff $x^*$ is important, and determines the exponent $\alpha$.

\subsection {Example III: the Fr\'{e}chet distribution}

The final example belongs to the third possible universality class, corresponding to variables with a power-law tail.

If at large $x$ we have:

\be p(x) = \frac{A_+}{(x-B)^{1+\mu}}, \ee

Then the cumulative distribution is:

\be C(x) = 1 - \frac{A_+}{\mu (x-B)^\mu}. \ee

Therefore taking it to a large power $n$ we find:

\be C^n(x) \approx e^{-\frac{A_+ n}{\mu (x-B)^\mu}}. \ee

Upon appropriately scaling the variable, we find that:

\be G(x) = e^{-a \left(\frac{x-b}{n^{1/\mu}}\right)^{-\mu}}, \label{frechet1}\ee
(where $a$ and $b$ do not depend on $n$).

Importantly, we see that in this case the width of the distribution increases with $n$ as a power-law $n^{1/\mu}$ -- for $\mu \leq 1$, this is precisely the same scaling we derived for the \emph{sum} of $n$ variables drawn from this heavy-tailed distribution! This elucidates why in the scenario $\mu \leq 1$ (corresponding to Fig. (\ref{sum_})c) we obtained \emph{L\'{e}vy flights}, where the sum was dominated by rare events no matter how large $n$ was. This is related to the so-called ``Single Big Jump Principle", which has been recently shown to pertain to a broader class of scenarios in physics, extending the results for i.i.d. variables (see Refs. \onlinecite{barkai1, barkai2} and references therein), as well as applications in finance \cite{filiasi2014concentration}.

We shall now show that these 3 cases can be derived in a general framework, using a similar approach to the one we used earlier.

\subsection {General form for Extreme Value Distributions}
We will now find all possible limiting distributions, following similar logic to the ``RG" used to find the form of the characteristic functions in the GCLT. By itself, our analysis will not reveal the ``basin of attraction" of each universality class, nor will we find the precise scaling of the coefficients $a_n$ and $b_n$. These require work beyond the basic RG calculation presented here.

As before, let us assume that there exists some scaling coefficients $a_n$, $b_n$ such that when we define:
\be \xi_n \equiv \frac{X_n-b_n}{a_n}, \label{EVD_scaling}\ee
the following limit exists:
\be \lim_{n \rightarrow \infty} Prob(\xi_n =\xi) = g(\xi). \ee
(note that this limit is not unique: we can always shift and rescale by a constant).
This would imply that $p(X_n) \approx a_n^{-1} g \left ( \frac{X_n-b_n}{a_n} \right )$ and the cumulative is given by: $G \left ( \frac{X_n-b_n}{a_n} \right )$.
By the same logic we used before, we know that $G^m \left ( \frac{X_n - b_n}{a_n} \right )$ depends only on the quantity $N=n \times m$. Therefore we have:
\be \frac{\partial}{\partial n} \left ( G^{N/n} \left ( \frac{X_n-b_n}{a_n} \right ) \right ) = 0. \ee
(Note that here $X_n$ is the random variable we are interested in: hence while derivatives of the \emph{coefficients} $a_n$, $b_n$ appear, a derivative of $X_n$ is not defined and does not appear).

%This leads to:
%\be \frac{\partial }{\partial n} \left (  \frac{N}{n} \log \left ( G \left ( \frac{X_n-b_n}{a_n} \right ) \right )\right)=0. \ee

From which we find:

\be - \frac{N}{n^2} \log G + \frac{N}{n} \frac{G'}{G} \left [ - \frac{\partial}{\partial n} \left ( \frac{b_n}{a_n} \right ) - \frac{X_n}{a_n^2} \frac{\partial a_n}{\partial n} \right ] = 0. \label{diff_}\ee

Upon defining a new random variable $\zz \equiv \frac{X_n-b_n}{a_n}$ (as in Eq. (\ref{EVD_scaling})), the equation can be rewritten as:

\be [\log (-\log G(\zz))]' = -\left [ \frac{n}{a_n} \frac{\partial b_n}{\partial n} +\frac{n}{a_n} \frac{\partial a_n}{\partial n}\zz \right ]^{-1}.\ee

In order for the RHS to have a sensible limit for large $n$, we would like to have:

\be \lim_{n \to \infty} \frac{n}{a_n} \frac{\partial a_n}{\partial n} = \alpha , \label{a_scaling} \ee
with $\alpha$ constant. Similarly, we have:

\be  \lim_{n \to \infty} \frac{n}{a_n} \frac{\partial b_n}{\partial n} = \beta , \label{b_scaling}\ee
with $\beta$ constant.

We shall shortly show that the value of $\alpha$ will dictate which of the three universality classes we will converge to.

{\bf Fr\'{e}chet Distribution:}\\

If $\alpha> 0$, we find that to leading order (with the same subtle interpretation as in the case of the L\'{e}vy stable distribution above):

\be  a_n \propto n^ \alpha. \ee

Next, requiring that $\frac{n}{a_n} \frac{\partial b_n}{\partial n}$ should be constant implies that $b_n \propto a_n$. This corresponds to a \emph{shift} in the scaled variable, and therefore we can set $b_n=0$ without loss of generality.

Solving for $G(\zz)$ gives the Fr\'{e}chet distribution:

\be G(x) = e^{-a x^{-1/\alpha}}. \label{frechet}\ee

Comparing this form with Eq. (\ref{frechet1}), we recognize that $1/\alpha = \mu$.

{\bf Weibull Distribution:}\\

Similarly, when $\alpha<0$ we find that to leading order:

\be  a_n \propto n^{-|\alpha|}. \ee

Solving for $b_n$ we find that to leading order it is a constant. Since this is \emph{not} the same scaling as $a_n$, in this case it does not correspond to a simple shift, and for this reason the choice of the constant is not arbitrary.
As expected, in order for the limit to exist we must choose $b_n = x^*$, where $x^*$ is the boundary of the (finite) support of the original variable.

Solving for $G$ gives us the Weibull Distribution:

\be G(x) = e^{a (x^*-x)^{1/\alpha}}. \label{weibull}\ee

As mentioned before, the coefficient $\alpha$ is determined by the behavior of the original probability distribution near the cutoff $x^*$. In the particular example discussed earlier $p(x)$ approached a non-zero constant near $x^*$, hence we found $\alpha=1$. It is straightforward to generalize this to the case where $p(x)$ vanishes near the cutoff as a power-law $(x-x^*)^c$, finding that $1/\alpha = c+1$.

{\bf Gumbel Distribution:}\\

Finally, consider the case $\alpha=0$. Given that $\frac{n}{a_n} \frac{\partial b_n}{\partial n}$ is approximately constant, we obtain the Gumbel distribution:

\be G(x)= e^{ -e^{- (a x +b)}}.\ee

In this case the RHS of Eq. (\ref{a_scaling}) vanishes. This implies that, unfortunately, the scaling coefficients $a_n$ cannot be determined -- not even to leading order -- and according to Eq. (\ref{b_scaling}) the same holds for the scaling coefficients $b_n$. Indeed, for $\alpha=0$ both scaling coefficients $a_n$ and $b_n$ are non-universal, and must be determined from the tail of $p(x)$. An interesting extended discussion can be found in Ref. \onlinecite{pierpaolo}, where the Gaussian case is analyzed. It can be shown that for the Gaussian distribution a particular (but non-unique) choice of scaling coefficients that leads to convergence to a Gumbel distribution is \cite{fisher, pierpaolo}:
\be a_n = 1/b_n; b_n = \sqrt{2 \log(n) - \log(4 \pi \log(n))}. \ee

\section{Summary}
The Generalized Central Limit Theorem and the Extreme Value Distribution are often referred to as tales of tails -- primarily dealing with distributions that are ``heavy-tailed", leading to the breakdown of the CLT. We began by exploring sums of (i.i.d.) random variables. We used a renormalization-group approach to find all possible limiting (stable) distributions of the sums, leading us to a generalization of the CLT to heavy-tailed distributions. Finally, we used a similar approach to study the similarly universal behavior of the \emph{maximum} of a large number of (i.i.d) variables, in which case the cumulative distribution played the part previously taken by the characteristic function. In both cases the self-similarity of the resulting sum or maximum led to a simple ODE governing the limiting distributions and elucidating its universal property. In the future, it would be interesting to see if this approach can be adapted to also yield the corrections to the leading order scaling and to the limiting distributions, as renormalization group approaches often do, as well as extend the approach to functions of multiple variables \cite{teuerle2012multidimensional}.

\textbf{Acknowledgments}
I thank Ori Hirschberg for numerous useful comments and for a critical reading of this manuscript. I thank the students of AM 203 (``Introduction to disordered systems and stochastic processes") at Harvard University where this material was first tested, as well as the participants of the Acre Summer School on Stochastic Processes with Applications to Physics and Biophysics held in September 2017, and in particular Eli Barkai. I thank Ethan Levien, Farshid Jafarpour and Pierpaolo Vivo for useful comments on the manuscript.

%\bibliography{GCLT}

\begin{thebibliography}{10}

\bibitem{feller}
Feller, W.
\newblock (2008) {\em An introduction to probability theory and its
  applications}.
\newblock (John Wiley \& Sons) Vol.{}~1.

\bibitem{CS}
Indyk, P.
\newblock (2000) {\em Stable distributions, pseudorandom generators, embeddings
  and data stream computation}.
\newblock (IEEE), pp. 189--197.

\bibitem{sethna}
Sethna, J.
\newblock (2006) {\em Statistical mechanics: entropy, order parameters, and
  complexity}.
\newblock (Oxford University Press) Vol.{}~14.

\bibitem{jona2001renormalization}
Jona-Lasinio, G.
\newblock (2001) Renormalization group and probability theory.
\newblock {\em Physics Reports} {\bf 352}, 439--458.

\bibitem{calvo2010generalized}
Calvo, I, Cuch{\'\i}, J.~C, Esteve, J.~G,  \& Falceto, F.
\newblock (2010) Generalized central limit theorem and renormalization group.
\newblock {\em Journal of Statistical Physics} {\bf 141}, 409--421.

\bibitem{taleb}
Taleb, N.~N.
\newblock (2007) {\em The black swan: The impact of the highly improbable}.
\newblock (Random house) Vol.{}~2.

\bibitem{fisher}
Fisher, R.~A \& Tippett, L. H.~C.
\newblock (1928) {\em Limiting forms of the frequency distribution of the
  largest or smallest member of a sample}.
\newblock (Cambridge University Press), Vol.{}~24, pp. 180--190.

\bibitem{RG1}
Bertin, E \& Gy{\"o}rgyi, G.
\newblock (2010) Renormalization flow in extreme value statistics.
\newblock {\em Journal of Statistical Mechanics: Theory and Experiment} {\bf
  2010}, P08022.

\bibitem{RG2}
Calvo, I, Cuch{\'\i}, J.~C, Esteve, J.~G,  \& Falceto, F.
\newblock (2012) Extreme-value distributions and renormalization group.
\newblock {\em Physical Review E} {\bf 86}, 041109.

\bibitem{sethna_PRL}
Manzato, C, Shekhawat, A, Nukala, P.~K, Alava, M.~J, Sethna, J.~P,  \& Zapperi,
  S.
\newblock (2012) Fracture strength of disordered media: Universality,
  interactions, and tail asymptotics.
\newblock {\em Physical review letters} {\bf 108}, 065504.

\bibitem{gyorgyi2008finite}
Gy{\"o}rgyi, G, Moloney, N, Ozog{\'a}ny, K,  \& R{\'a}cz, Z.
\newblock (2008) Finite-size scaling in extreme statistics.
\newblock {\em Physical review letters} {\bf 100}, 210601.

\bibitem{gyorgyi2010renormalization}
Gy{\"o}rgyi, G, Moloney, N, Ozog{\'a}ny, K, R{\'a}cz, Z,  \& Droz, M.
\newblock (2010) Renormalization-group theory for finite-size scaling in
  extreme statistics.
\newblock {\em Physical Review E} {\bf 81}, 041135.

\bibitem{barkai1}
Vezzani, A, Barkai, E,  \& Burioni, R.
\newblock (2019) Single-big-jump principle in physical modeling.
\newblock {\em Physical Review E} {\bf 100}, 012108.

\bibitem{barkai2}
Wang, W, Vezzani, A, Burioni, R,  \& Barkai, E.
\newblock (2019) Transport in disordered systems: the single big jump approach.
\newblock {\em arXiv preprint arXiv:1906.04249}.

\bibitem{filiasi2014concentration}
Filiasi, M, Livan, G, Marsili, M, Peressi, M, Vesselli, E,  \& Zarinelli, E.
\newblock (2014) On the concentration of large deviations for fat tailed
  distributions, with application to financial data.
\newblock {\em Journal of Statistical Mechanics: Theory and Experiment} {\bf
  2014}, P09030.

\bibitem{pierpaolo}
Vivo, P.
\newblock (2015) Large deviations of the maximum of independent and identically
  distributed random variables.
\newblock {\em European Journal of Physics} {\bf 36}, 055037.

\bibitem{teuerle2012multidimensional}
Teuerle, M, {\.Z}ebrowski, P,  \& Magdziarz, M.
\newblock (2012) Multidimensional levy walk and its scaling limits.
\newblock {\em Journal of Physics A: Mathematical and Theoretical} {\bf 45},
  385002.

\end{thebibliography}
%\bibliographystyle{pnas2009}

\newpage

%\section{Appendix: Code for running sum}
\appendix{\textbf{Appendix: Code for running sum}}

In order to generate the data shown in Fig. \ref{sum_}(a), the following MATLAB code is used:

\begin{verbatim}
tmp=randn(N,1);
x=cumsum(tmp);
\end{verbatim}

For Fig. \ref{sum_}(b), to generate a running sum of variables drawn from the Cauchy distribution, the first line is replaced with:

\begin{verbatim}
tmp = tan(pi*(rand(N,1)-1/2));
\end{verbatim}

Finally, for Fig. \ref{sum_}(c) the same line is replaced with:

\begin{verbatim}
t=rand(N,1);
b=1/mu;
tmp=t.^(-b);
\end{verbatim}

where we used $\mu=1/2$ for the figure.

\end{document}